\documentclass[apj, twocolumn, numberedappendix, appendixfloats, twocolappendix]{emulateapj}

\newcommand{\myemail}{yutaka.hirai@nao.ac.jp}
\renewcommand{\thefootnote}{\fnsymbol{footnote}}
\shorttitle{Enrichment of $r$-process elements in dSphs}
\shortauthors{Hirai et al.}
\begin{document}
\title{Enrichment of $r$-process elements in dwarf spheroidal galaxies in chemo-dynamical evolution model}
\author{Yutaka Hirai\altaffilmark{1,2,7}}
\author{Yuhri Ishimaru\altaffilmark{3, 4}}
\author{Takayuki R. Saitoh\altaffilmark{5}}
\author{Michiko S. Fujii\altaffilmark{2}}
\author{Jun Hidaka\altaffilmark{6,2}}
\author{\\Toshitaka Kajino\altaffilmark{2,1}}
\affil{\altaffilmark{1}Department of Astronomy, Graduate School of Science, the University of Tokyo, 7-3-1 Hongo, Bunkyo-ku, Tokyo 113-0033, Japan\\\altaffilmark{2}Division of Theoretical Astronomy, National Astronomical Observatory of Japan, 2-21-1 Osawa Mitaka, Tokyo 181-8588, Japan; \myemail\\
\altaffilmark{3}Department of Material Science, International Christian University, 3-10-2 Osawa, Mitaka, Tokyo 181-8585, Japan\\
\altaffilmark{4}Institut d'Astrophysique de Paris, 
98bis Boulevard Arago, 75014, Paris, France\\
\altaffilmark{5}Earth-Life Science Institute, Tokyo Institute of Technology, 2-12-1 Ookayama, Meguro-ku, Tokyo 152-8551, Japan\\
\altaffilmark{6}School of Science and Engineering, Meisei University, 2-1-1 Hodokubo, Hino, Tokyo 191-0042, Japan}
\altaffiltext{7}{Research Fellow of Japan Society for the Promotion of Science}
\begin{abstract}
The rapid neutron-capture process ($r$-process) is a major process to synthesize elements heavier than iron, but the astrophysical site(s) of $r$-process is not identified yet. Neutron star mergers (NSMs) are suggested to be a major $r$-process site from nucleosynthesis studies. Previous chemical evolution studies however require unlikely short merger time of NSMs to reproduce the observed large star-to-star scatters in the abundance ratios of $r$-process elements relative to iron, [Eu/Fe], of extremely metal-poor stars in the Milky Way (MW) halo. This problem can be solved by considering chemical evolution in dwarf spheroidal galaxies (dSphs) which would be building blocks of the MW and have lower star formation efficiencies than the MW halo. We demonstrate that enrichment of $r$-process elements in dSphs by NSMs using an $N$-body/smoothed particle hydrodynamics code. Our high-resolution model reproduces the observed [Eu/Fe] by NSMs with a merger time of 100 Myr when the effect of metal mixing is taken into account. This is because metallicity is not correlated with time up to $\sim$ 300 Myr from the start of the simulation due to low star formation efficiency in dSphs. We also confirm that this model is consistent with observed properties of dSphs such as radial profiles and metallicity distribution. The merger time and the Galactic rate of NSMs are suggested to be $\lesssim$ 300 Myr and $\sim 10^{-4}$ yr$^{-1}$, which are consistent with the values suggested by population synthesis and nucleosynthesis studies. This study supports that NSMs are the major astrophysical site of $r$-process. 
\end{abstract}
\section{Introduction}
Elements heavier than iron are mainly synthesized by the rapid neutron capture process ($r$-process) as well as the slow neutron capture and proton capture processes. More than 90\% of the number of elements such as europium (Eu), gold (Au), and platinum (Pt)\footnote{Hereafter, we call these elements $r$-process elements} in the solar system are synthesized by $r$-process \citep{Burris00}. Sufficiently neutron-rich environment is required in order to synthesize $r$-process elements with mass number ($A$) over 110.\\
\indent The observed $r$-process elemental abundance ratios, such as [Eu/Fe]\footnote{[A/B] = $\log_{10}({N_{\mathrm{A}}}/{N_{\mathrm{B}}})-\log_{10}({N_{\mathrm{A}}}/{N_{\mathrm{B}}})_{\odot}$, where $N_{\mathrm{A}}$ and $N_{\mathrm{B}}$ are number densities of elements A and B, respectively. }, in extremely metal-poor (EMP) stars ([Fe/H] $\lesssim -$3) show large star-to-star scatters. These scatters are seen in EMP stars in the Milky Way (MW) halo as well as the Local Group (LG) dwarf spheroidal galaxies (dSphs)  \citep[e.g., ][]{Woolf95, McWilliam95, Ryan96, Shetrone96, McWilliam98, Westin00, Burris00, Fulbright00, Norris01, Johnson02, Francois03, Honda04, Francois07, Sneden08, Tolstoy09, Frebel10a, Frebel10b, Letarte10, Aoki13, Ishigaki13}. The mechanism of star-to-star scatters in the abundance studies should be clarified simultaneously with the astrophysical site(s) of $r$-process.\\
\indent Neutrino-driven winds from proto-neutron stars of core-collapse supernovae (CCSNe) have long been regarded as one of the possible sites of $r$-process \citep[e.g., ][]{Meyer92, Woosley94, Qian96, Wanajo01}. Previous chemical evolution studies suggest that the observed [Eu/Fe] scatter is well reproduced by models assuming that CCSNe of low-mass (8--10 $M_{\odot}$) progenitors produce $r$-process elements \citep[e.g., ][]{Mathews92, Ishimaru99, Travaglio99, Tsujimoto00, Travaglio01, Ishimaru04, Argast04}. However, recent hydrodynamical simulations of CCSNe, which include neutrino transport in sophisticated manner, suggest that the PNS winds of CCSNe do not necessarily produce a neutron-rich condition suitable for $r$-process \citep[e.g., ][]{Reddy98, Roberts12, Martinez12, Roberts+12, Horowitz12}. Nucleosynthesis calculations suggest that heavy elements with $A \gtrsim$ 110, are difficult to be synthesized in CCSNe due to such too weak neutron-rich environments \citep[e.g., ][]{Wanajo11, Wanajo13}.\\ 
\indent Binary neutron star mergers (NSMs) are also suggested to be a promising site of $r$-process \citep{Lattimer74, Lattimer76, Lattimer77, Symbalisty82, Eichler89, Meyer89}. Recent detailed nucleosynthesis calculations show that heavy $r$-process elements are successfully synthesized in NSMs \citep{Freiburghaus99, Goriely11, Korobkin12, Bauswein13, Rosswog14, Wanajo14}. In addition, near infrared afterglow of the {\it Swift} short gamma-ray burst, GRB130603B \citep{Berger13, Tanvir13} is detected. This is suggested to be a piece of evidence that progenitors of short gamma-ray bursts are compact binary mergers and $r$-process nucleosynthesis occurs there \citep{Tanaka13}.\\
\indent NSMs have the long merger time and the low rate. The neutron star (NS) binaries lose their energy very slowly due to the gravitational emissions. The merger time is estimated $\gtrsim$ 100 Myr from observed binary pulsars \citep{Lorimer08}. Recent predictions from population synthesis models also suggest that most of NS binaries merge $\gtrsim$ 100 Myr after their formation \citep{Dominik12}. On the other hand, the NSM rate is estimated to be $10^{-6}$ -- $10^{-3}$ yr$^{-1}$ for a MW size galaxy from the observed binary pulsars \citep{Abadie10}.\\
\indent However, early studies of galactic chemical evolution pointed out that it is difficult to reproduce the observed trend of [Eu/Fe] of EMP stars by NSMs due to their long merger time and the low rate \citep{Mathews90, Argast04}. Most of other recent studies also conclude that short merger time ($\lesssim$ 10 Myr) or a second site of $r$-process such as jet-like explosions of magnetorotational CCSNe \citep[e.g., ][]{Winteler12, Nishimura15} is required to account for large star-to-star scatters in EMP stars \citep{Matteucci14, Komiya14, Tsujimoto14, Cescutti15, Wehmeyer15}. On the other hand, detailed population synthesis calculations suggest that production of NSMs with short-merger time ($\sim$ 10 Myr) highly depends on the treatment of common envelop phase which is not well understood \citep[e.g.,][]{Portegies98, Dominik12, Kinugawa14}. \citet{Dominik12} suggest that NSMs with merger time of $\lesssim$ 10 Myr cannot be produced in their most pessimistic model assuming each common envelop with an Hertzsprung gap donor causes a merger \citep[Submodel B in][]{Dominik12}. In addition, there are no observational clues that exist binary pulsars which merge within $\sim$ 10 Myr so far \citep{Lorimer08}.\\
\indent This discrepancy may be solved, if the Galactic halo is formed via mergers of sub-halos within the framework of hierarchical structure formation scenario \citep{Ishimaru14}. \citet{Ishimaru14} calculate the enrichment of $r$-process elements by NSMs with merger time of 100 Myr (95 \% of NSMs) and 1 Myr (5 \% of NSMs) in their one-zone chemical evolution model for each sub-halo. They suggest that [Eu/Fe] increases at [Fe/H] $\leq -3$ if the star formation efficiencies are lower in less massive sub-halos. According to their calculations, the observed scatters in [Eu/Fe] in metal-poor stars are possibly explained by NSMs with merger time of 100 Myr. Key factors of chemical evolution such as the time variation of the star formation rate (SFR), outflow, and inflow strongly depend on thermodynamical feedback from SNe. Detailed chemo-dynamical evolution studies in low-mass galaxies like sub-halos are highly desirable to justify their assumptions to describe the enrichment history of $r$-process elements in a self-consistent manner between dynamical and chemical evolution of galaxies.\\
\indent Recent hydrodynamical studies have performed a series of simulations of galaxy formation assuming that the NSMs are a major site of $r$-process \citep{Voort15, Shen15}. \citet{Voort15} suggest that gas mixing processes such as galactic winds and hydrodynamic flows play the important roles to explain the observed Galactic $r$-process ratio. Their high-resolution model with $7.1\times10^{3}M_{\odot}$ of the mass of one gas particle ($m_{\mathrm{gas}}$) is, however, difficult to reproduce the observed $r$-process abundance ratios. They imply that additional metal mixing is required to explain the observation. \citet{Shen15} also suggest that the observed $r$-process abundance ratios are possibly taken into account in their NSM models if the metal mixing in star-forming region is implemented. However, the mass resolution ($m_{\mathrm{gas}} = 2.0\times10^{4} M_{\odot}$) in their models is as low as the fiducial low-resolution model of \citet{Voort15} ($m_{\mathrm{gas}} = 5.7\times10^{4} M_{\odot}$). It is therefore important to demonstrate that if the NSM models account for the observations in independent simulations of much higher resolution.\\
\indent In this paper, we calculate enrichment of $r$-process elements ejected by NSMs in low mass galaxies with high-resolution ($m_{\mathrm{gas}} = 4.0\times10^{2} M_{\odot}$) chemo-dynamical evolution models. We perform $N$-body/smoothed particle hydrodynamics (SPH) simulations of dSph models using ASURA code \citep{Saitoh08, Saitoh09}. We discuss effects of metal mixing in star-forming region as well as the dependence on the SFR, the merger time, and the rate of NSMs.\\
\indent In \S 2, we introduce our code and models. In \S 3, we compare model predictions and observed properties of dSphs generated by our models. In \S 4, we discuss enrichment of $r$-process elements in dSphs. Finally in \S 5, we summarize our main results.
\section{Method and Models}
\subsection{$N$-body / smoothed particle hydrodynamics code, ASURA}\label{ASURA}
We perform a series of simulations using an $N$-body / SPH code, ASURA \citep{Saitoh08, Saitoh09}. We adopt three different kinds of particles in our simulations: dark matter, gas, and star particles. We treat dark matter and star particles as collisionless particles. Dark matter particles contribute to the dynamical evolution of our model galaxies. Star particles mainly contribute to feedback of energy and heavy elements produced by CCSNe. We solve the hydrodynamical evolution of gas particles using SPH \citep[e.g., ][]{Lucy77, Gingold77, Monaghan&Lattanzio85,Monaghan92}. \\
\indent Here we describe implementation of gravity and hydrodynamics in ASURA. Gravity is calculated by a treecode \citep{Barnes86} with GRAPE method \citep{Makino91} using the Phantom-GRAPE library \citep{Tanikawa12}. Hydrodynamics is computed using a standard SPH method. For integration of self-gravitating fluid system, we adopt the fully asynchronous split time-integrator (FAST) algorithm in order to reduce calculation cost \citep{Saitoh&Makino10}. We use the time-step limiter, which forces the timestep difference among neighbor particles to be less than four times long, in order to follow the evolution of strong shock regions such as supernova (SN) remnants \citep{Saitoh&Makino09}.  We use a metallicity dependent cooling/heating function generated by Cloudy \citep{Ferland98, Ferland13}. The cooling/heating function covers the temperature range from 10 K to 10$^{9}$ K.\\
\indent We allow star formation when gas particles satisfy three conditions. (1) Gas particles are conversing ($\nabla\cdot\bf{v} <$ 0). (2) The density is higher than threshold density, $n_{\mathrm{th}}$. (3) The temperature is lower than threshold temperature, $T_{\mathrm{th}}$ \citep[e.g., ][]{Navarro&White93, Katz96, Stinson06}. We adopt $n_{\mathrm{th}}$ = 100 cm$^{-3}$ for our fiducial model, which is the mean density of giant molecular clouds (GMCs). We adopt $T_{\mathrm{th}}$ = 1000 K for our fiducial model. The value of $T_{\mathrm{th}}$ is insensitive to the final structure of galaxies \citep{Saitoh08}.\\
\indent When a gas particle satisfies the three conditions above, it becomes eligible to form new collisionless star particles. Star particles are produced by the following probability according to the prescription of \citet{Katz92, Katz96}:
\begin{equation}
p = \frac{m_{\mathrm{gas}}}{m_{\star}}
\left\{1-\exp\left(-c_{\star}\frac{dt}{t_{\mathrm{dyn}}}\right)\right\},
\end{equation}
where $m_{\star}$ and $m_{\mathrm{gas}}$ are the mass of star and gas particles, respectively, and $c_{\star}$ is the dimensionless star formation efficiency (SFE) parameter. We set $m_{\star} = m_{\mathrm{gas}}/3$ following \citet{Okamoto03, Okamoto05}. The mass of a gas particle of our fiducial model is initially assumed as $4.0 \times 10^{2} M_{\odot}$, while it is reduced by star formation. When the mass of a gas particle becomes lower than one-third of initial mass, the particle is converted into a collisionless particle. The dimensionless SFE parameter of our fiducial model ($c_{\star}$ = 0.033) is chosen based on the slow star-formation model \citep{Zuckerman74, Krumholz07}. 
\citet{Saitoh08} suggest that when $n_{\mathrm{th}}$ = 100 cm$^{-3}$ is adopted, the final results are fairly insensitive to the adopted value of $c_{\star}$ in their MW model. In the Appendix, we also confirm this result in our dSph models.\\
\indent Each star particle is treated as a single stellar population (SSP), i.e. each star particle is assumed to be an assembly of stars with the same age and the same metallicity. The initial mass function (IMF) of star particles is the Salpeter IMF: $\phi\equiv m^{-x}$, where $x = 1.35$  \citep{Salpeter55}  with mass range of 0.1--100$M_{\odot}$. We set the progenitor mass of CCSNe to be 8--40 $M_{\odot}$. In this model, stars more massive than 40 $M_{\odot}$ end their lives as black holes. Star particles, which explode in a time interval of $\Delta t$ are selected by the following probability ($p_{\mathrm{CCSNe}}$), 
\begin{equation}
p_{\mathrm{CCSNe}} = \frac{\displaystyle
\int^{m(t+\Delta t)}_{m(t)}
\phi(m^{\prime})m^{\prime -1}
\mathrm{d}m^{\prime}}{\displaystyle\int^{8 M_{\odot}}_{m(t)}\phi(m^{\prime})
m^{\prime -1}
\mathrm{d}m^{\prime}},
\end{equation}
where $m (t)$ is the turn off mass at age $t$. Each CCSN explosion distributes thermal energy of $10^{51}$ erg to the surrounding SPH particles. The mass of one star particle is $\sim 100 M_{\odot}$. When a particle explode as CCSNe, the number of CCSNe inside each star particle corresponds to $\sim 1$. This method is also adopted in other studies \citep[e.g., ][]{Okamoto08, Saitoh08}. In addition to the SN feedback, we implement heating by H$_{\mathrm{II}}$ region formed around young stars. The number of the Ly$\alpha$ photons is evaluated using P\'EGASE \citep{Fioc97}. 
The parameters of these baryonic physics are listed in Table \ref{baryon}.
\begin{deluxetable*}{llrr}[htbp]
\tabletypesize{\scriptsize}
\tablecaption{Parameters of baryon physics.\label{baryon}}
\tablewidth{0pt}
\tablehead{
\colhead{Quantity} & \colhead{Symbol} & \colhead{Fiducial values\tablenotemark{a}} & \colhead{Variation}
}
\startdata
		Dimensionless SFE parameter&$c_{\star}$&0.033&0.033, 0.5\\
		Threshold density for star formation&$n_{\mathrm{th}}$&100 cm$^{-3}$&0.1--100 cm$^{-3}$\\
		Threshold temperature for star formation&$T_{\mathrm{th}}$&1$\times$10$^{3}$ K&1$\times$10$^{3}$ -- 3$\times$10$^{4}$ K \\
	         SN explosion energy&$\epsilon_{\mathrm{SN}}$&1 $\times$10$^{51}$ erg &(0.03 -- 1) $\times$10$^{51}$erg
\enddata
\tablenotetext{a}{Fiducial values of $c_{\star}$, $n_{\mathrm{th}}$, $T_{\mathrm{th}}$, $\epsilon_{\mathrm{SN}}$ are taken from \cite{Saitoh08}.} 
\end{deluxetable*}
\subsection{Chemical enrichment process} \label{Chem}
We take into account both CCSNe and NSMs in our models. We set initial gas metallicity equals to zero. CCSNe produce Fe and NSMs produce Eu, which is regarded as a representative element of $r$-process. Binary black hole-neutron star mergers are also expected to eject $r$-process elements \citep[e.g., ][]{Korobkin12, Mennekens14}. However, they affect the rate of production of $r$-process elements by several factors which are much smaller than the uncertainty of the rate of NSMs. We therefore only implement NSMs for simplicity. We assume that gas particles around a star particle are enriched with metals when a CCSN or NSM occurs in a star particle. Metals are distributed in 32 nearest neighbor particles using weights of SPH kernel. Mass of element X in the $j$th neighbor particle ejected by $i$th star particle, $\Delta M_{\mathrm{X},j}$, is given by
\begin{equation}\label{metal}
\Delta M_{\mathrm{X},j} = \frac{m_{j}}{\rho_{i}}M_{\mathrm{X},i}W(r_{ij}, h_{ij}),
\end{equation}
 where  $r_{ij}$ is the distance between particle $i$ and $j$, $h_{ij}$ is the smoothing length, and $W(r_{ij}, h_{ij})$ is the SPH kernel given by a cubic spline function \citep[e.g., ][]{Kawata01} and the density of the gas particles is given as
 \begin{equation}
 \rho_{i} = \sum_{i\neq j}m_{j}W(r_{ij}, h_{ij}).
 \end{equation} 
Smoothing length is the scale of containing $N_{\mathrm{ngb}}$ nearest neighbor particles. Following \cite{Saitoh08}, we set $N_{\mathrm{ngb}}$ = 32$\pm$2 as a fiducial value.\\
\indent The iron yield of CCSNe are taken from \citet{Nomoto06}. The scope of this paper is to discuss the abundance ratio of $r$-process elements in EMP stars before type Ia supernovae (SNe Ia) contribute. We do not implement SNe Ia in our simulation.\\
\indent The NSM rate and the merger time ($t_{\mathrm{NSM}}$) are highly uncertain. We therefore vary them $\sim$ 2 dex in our simulations. We regard a number fraction of NSMs to the total number of neutron stars, $f_{\mathrm{NSM}}$, as a parameter, which determines the NSM rate. In this model, we assume the mass range of NS progenitor mass as  8--20 $M_{\odot}$. We set the upper mass of NSM progenitor stars as 20 $M_{\odot}$ from the lower limit of the mass of a black hole formation \citep{Dominik12}. We set $f_{\mathrm{NSM}}$ = 0.01 as a fiducial value. The corresponding NSM rate in a Milky-Way size galaxy is $\sim$ 10$^{-4}$ yr$^{-1}$. It is within the values of the Galactic disk $\sim$10$^{-6}$--10$^{-3}$ yr$^{-1}$, estimated from observed compact binaries \citep{Abadie10}. \\
\indent Yields of NSMs are related to the rate of NSMs. [Eu/Fe] at [Fe/H] = 0 is expected to be $\sim$ 0.5 without SNe Ia because solar Fe is estimated to be produced $\sim$60--65\%  by SNe Ia, and $\sim$35--40\% by CCSNe \citep[e.g., ][]{Goswami00, Prantzos08}.  We thus simply set the yield of $r$-process elements to be [Eu/Fe] = 0.5 at [Fe/H] = 0.\\
\indent Observed $r$-process elemental abundance ratio such as [Eu/Fe] indicates that the production of $r$-process elements should have occurred before SNe Ia start to contribute ($\gtrsim$ 1 Gyr) to galactic chemical evolution \citep{Maoz12}. Minimum merger time of NSMs needs to be shorter than the typical delay time of SNe Ia. As already mentioned, the most plausible merger time of NSMs is regarded as $\sim$ 100 Myr \citep[e.g., ][]{Lorimer08, Dominik12}. We thus set the merger time of NSMs is 100 Myr as a fiducial value. 
\subsection{Definition of abundance of newly formed stars}\label{metal-stars}
The abundance of a star must be identical to the abundance of the gas, which formed the star. The abundance of a newly formed star particle inherits (1) the abundance of the star-forming gas particle \citep[e.g., ][]{Raiteri99, Voort15, Shen15}; (2) the abundance of the average of gas particles within a SPH kernel \citep[e.g., ][]{Steinmetz94, Kobayashi11, Shen15}. Method (1) does not have a metal mixing process except for hydrodynamical mixing process such as stellar winds, outflows and inflows due to the SN explosion. [Eu/Fe] produced in method (1) is discussed in \S\ref{mix1}. On the other hand, in method (2), a metal mixing process is taken into account. We use a metallicity averaged over 32 neighbor gas particles in a SPH kernel to a newly born star particle. The region can be regarded as star-forming region, which corresponds to $\sim 10^4 M_{\odot}$. This mass corresponds to the typical size of giant molecular clouds (GMCs) \citep[e.g., ][]{Larson81, Liszt81, Sanders85, Solomon87, William94, Heyer09}. We discuss results inferred from method (2) in \S\ref{mix2}.\\
\indent Massive stars tend to be born in clusters and associations \citep{Lada03}. Clusters and OB associations form from GMCs. Observations of stars in open clusters suggest that their metallicity is homogeneous \citep{De Silva07a, De Silva07b, Pancino10, Bubar10, De Silva11, Ting12, Reddy12, De Silva13, Reddy13}. \citet{Feng14} theoretically show that turbulent mixing in star-forming regions causes this homogeneity. The timescale of metal mixing is determined by the local dynamical time of star-forming regions ($\lesssim$ 1 Myr). This timescale is much shorter than the typical timescale of star formation ($\gtrsim$10 Myr) in slow star formation model \citep{Zuckerman74, Krumholz07}. We thus assume that the metals are instantaneously mixed in star-forming regions. We discuss the effect of different implementation on abundance ratios in galaxies in \S\ref{mix1} and \ref{mix2}.
\subsection{Models of dSphs}
We follow initial conditions of dSph models adopted in \citet{Revaz09} and \citet{Revaz12}. We assume the density profile of dark matter, $\rho$, as follows:
\begin{equation}\label{pseudo}
\rho=\frac{\rho_c}{1+(r/r_c)^2},
\end{equation}
where $\rho_c$ is the central density and $r_c$ is the core radius. Gas particles are also distributed along with the profile. We set $r_{\mathrm{c}}$ = 1.0 kpc, $r_{\mathrm{max}}$ = 7.1 kpc, and $M_{\mathrm{tot}}$ =  7 $\times 10^{8} M_{\odot}$, according to \citet{Revaz12}. We adopt 0.15 of mass ratio of gas to dark matter particles (baryon fraction, $f_{\mathrm{b}} \equiv\Omega_{\mathrm{b}} / \Omega_{\mathrm{m}}$). The value of $f_{\mathrm{b}}$ is taken from \citet{Planck13}.\\
\indent Following \citet{Revaz09}, we assume an isotropic velocity dispersion of dark matter particles, $\sigma(r)$, for spherical distribution \citep{Hernquist93, Binney08},
\begin{equation}
\sigma^{2}(r) = \frac{1}{\rho(r)}\int^{\infty}_{r}\mathrm{d}r^{\prime}\rho(r^{\prime})\frac{\partial\Phi(r^{\prime})}{\partial{r^{\prime}}},
\end{equation}
where $\Phi(r)$ is the gravitational potential. For gas particles, we set a velocity equal to zero and an initial temperature of $10^{4}$ K.  For both dark matter and gas particles, we adopt gravitational softening length ($\epsilon_{\mathrm{g}}$) of 28 pc for runs with the initial total number ($N$) of $2^{14}$, $\epsilon_{\mathrm{g}}$ = 14 pc for runs with $N = 2^{16}$, $2^{17}$, and $2^{18}$, and $\epsilon_{\mathrm{g}}$ = 7 pc for runs with $N$ = $2^{19}$. We run our simulations over 14 Gyr. The parameters of our model galaxies are listed in Table \ref{model}. Table \ref{list} summarizes all runs discussed in this paper. 
\begin{deluxetable*}{lll}[htbp]
\tabletypesize{\scriptsize}
\tablecaption{Parameters of the initial condition.\label{model}}
\tablewidth{0pt}
\tablehead{
\colhead{Quantity} & \colhead{Symbol} & \colhead{Values\tablenotemark{a}}}
\startdata
		Total mass&$M_{\mathrm{tot}}$&7$\times$10$^{8} M_{\odot}$\\
		Core radius&$r_{\mathrm{c}}$&1 kpc\\
		Initial outer radius&$r_{\mathrm{max}}$&7.1 kpc\\
		Baryon fraction&$f_{\mathrm{b}}$&0.15
\enddata
\tablenotetext{a}{Values are taken from \cite{Revaz09, Revaz12}.}
\end{deluxetable*}
\begin{deluxetable*}{llrrrcrllcrl}
\tabletypesize{\scriptsize}
\tablecaption{List of models. \label{list}}
\tablewidth{0pt}
\tablehead{
\colhead{Model}&
\colhead{$N$}&
\colhead{$m_{\mathrm{DM}}$}&
 \colhead{$m_{\mathrm{gas}}$}&
\colhead{$\epsilon_{\mathrm{g}}$}&
\colhead{$n_{\mathrm{th}}$}&
\colhead{$T_{\mathrm{th}}$}&
\colhead{$\epsilon_{\mathrm{SN}}$}&
\colhead{N$_\mathrm{ngb}$}&
\colhead{Mixing}&
\colhead{$t_{\mathrm{NSM}}$}&
\colhead{$f_{\mathrm{NSM}}$}\\
	&&(10$^{3}M_{\odot}$)&(10$^{3}M_{\odot}$)&\colhead{(pc)}&\colhead{(cm$^{-3}$)}&($10^{3}$K)&\colhead{(10$^{51}$ erg)}&&&\colhead{(Myr)}&}
\startdata
		s000&$2^{19}$&2.3&0.4&7&100&1&1&32&no&100&0.01\\\tableline			
		m000&$2^{19}$&2.3&0.4&7&100&1&1&32&yes&100&0.01\\
		mN16&$2^{19}$&2.3&0.4&7&100&1&1&16&yes&100&0.01\\
		mN64&$2^{19}$&2.3&0.4&7&100&1&1&64&yes&100&0.01\\
		m014&$2^{14}$&72.6&12.8&28&100&1&1&32&yes&100&0.01\\
		m016&$2^{16}$&18.2&3.2&14&100&1&1&32&yes&100&0.01\\
		m017&$2^{17}$&9.1&1.6&14&100&1&1&32&yes&100&0.01\\
		m018&$2^{18}$&4.5&0.8&14&100&1&1&32&yes&100&0.01\\\tableline
		mExt&$2^{19}$&2.3&0.4&7&0.1&30&0.03&32&yes&100&0.01\\\tableline
		mt10&$2^{19}$&2.3&0.4&7&100&1&1&32&yes&10&0.01\\
		mt500&$2^{19}$&2.3&0.4&7&100&1&1&32&yes&500&0.01\\\tableline
		mr0.001&$2^{19}$&2.3&0.4&7&100&1&1&32&yes&100&0.001\\
		mr0.1&$2^{19}$&2.3&0.4&7&100&1&1&32&yes&100&0.1
	\enddata
\tablecomments{Parameters adopted in our models:
(1)Model: Name of our models. Models named ``000" adopt the fiducial parameter set. Model s000 is discussed in \S\ref{chemodynamics} and \S\ref{mix1}. Models m000 to m018 are discussed in \S \ref{mix2}. Model mExt is discussed in \S\ref{p_m}. Models mt10 and mt500 are discussed in \S\ref{tdel}. Models mr0.001 and mr0.1 are discussed in \S\ref{NSMrate}. (2) $N$: Initial total number of particles. (3) $m_{\mathrm{DM}}$: Mass of one dark matter particle. (4) $m_{\mathrm{gas}}$: Initial mass of one gas particle. (5) $\epsilon_{\mathrm{g}}$: Gravitational softening length. (6) $n_{\mathrm{th}}$: Threshold density for star formation. (7) $T_{\mathrm{th}}$: Threshold temperature for star formation. (8) $e_{\mathrm{SN}}$: SN feedback energy. (9) $N_{\mathrm{ngb}}$: Number of nearest neighbor particles. (10) Mixing:  With (yes) or without (no) metal mixing in star-forming region. (11) $t_{\mathrm{NSM}}$: Merger time of NSMs. (12) $f_{\mathrm{NSM}}$: Fraction of NSMs.}
\end{deluxetable*}
\section{Chemo-dynamical evolution of dwarf spheroidal galaxies}\label{chemodynamics}
\subsection{Dynamical evolution of dSph models}\label{property}
\indent We discuss chemo-dynamical evolution of model s000 to confirm that the parameter set of this model is appropriate for the case of dSphs. Parameter dependence is discussed in the Appendix.\\
\indent Figure \ref{gas} shows the evolution of the spatial distribution of gas and stars of model s000. Upper panels of Figure \ref{gas} show the gas density maps of model s000 at 0 Gyr, 1 Gyr, 5 Gyr, and 10 Gyr from the beginning of the simulation. The gas monotonically collapses during the first 1 Gyr. Then, gas density reduces by star formation and gas outflow by energy feedback of SNe. Red colored area in upper panels of Figure \ref{gas} corresponds to the star-forming region where the number density of gas is larger than 100 cm$^{-3}$. As shown in this figure, star-forming region is strongly confined at the center of the galaxy. The red area is largest at 1 Gyr. This indicates that star formation is most active after $\sim$ 1 Gyr from the beginning of the simulation.\\
\indent Lower panels of Figure \ref{gas} show stellar density maps at 0 Gyr, 1 Gyr, 5 Gyr, and 10 Gyr. The distribution of stars at 1 Gyr is associated with the high-density region of gas (red region in Figure \ref{gas}). As shown in these figures, stars continuously form in the inner region of this model galaxy for over 10 Gyr and stellar density distribution expands with time from the center to the outer region.  At 10 Gyr, the morphology of the galaxy becomes spherical symmetry.\\
\begin{figure*}[htbp]
\epsscale{1.3}
\plotone{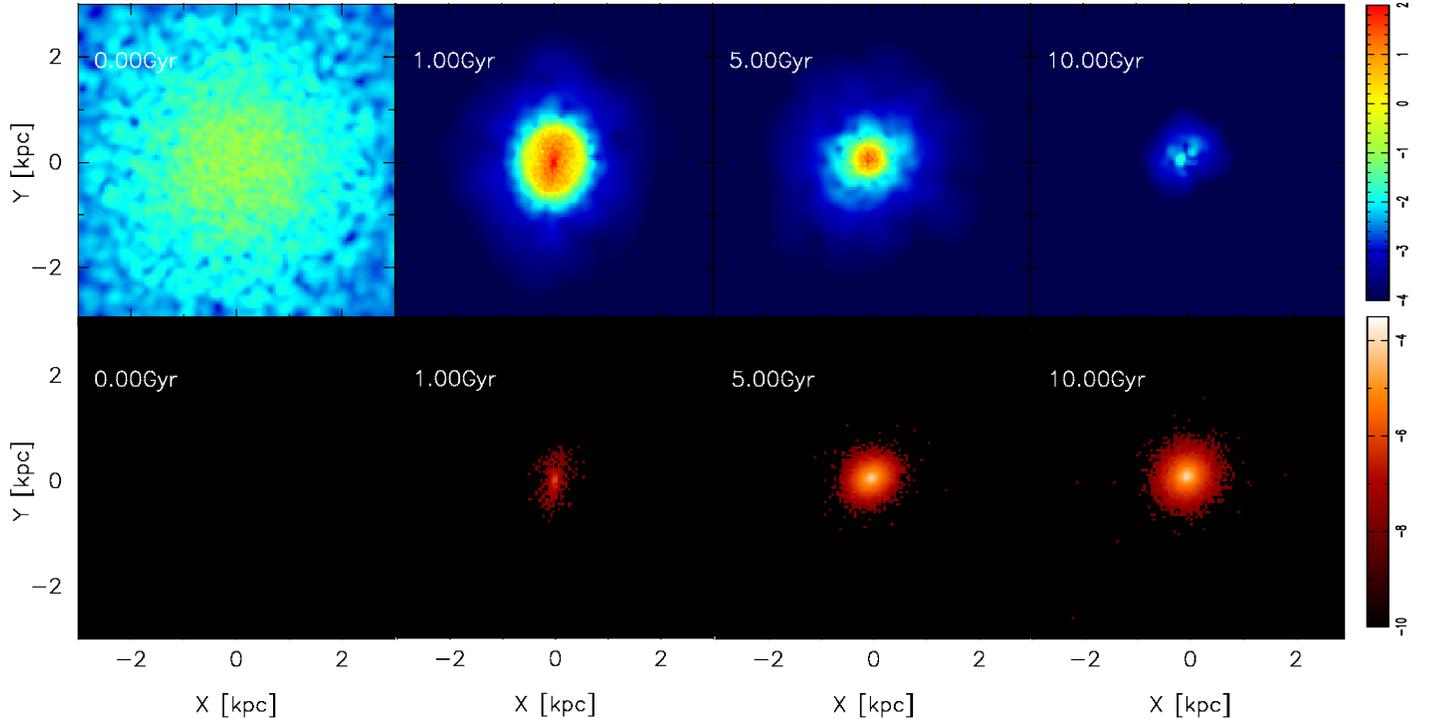}
\caption{Evolution of model s000. Upper panels: snapshots of slice gas density in log scale, between 10$^{-4}$ cm$^{-3}$ (blue) and 10$^2$ cm$^{-3}$ (red). Lower panels: snapshots of stellar surface density with log scale, between $10^{-10} 10^{10}M_{\odot}$kpc$^{-2}$ (black) and 10$^{-3.5} 10^{10}M_{\odot}$kpc$^{-2}$ (white).\label{gas}}
\end{figure*}
In order to quantitatively discuss the structural and dynamical properties of models, we investigate the radial profiles. Figure \ref{RP} shows the radial profiles of model s000. We define the galactic center using potential minimum. The values in each point are calculated in each bin from the center to the outer region.\\
\indent Figure \ref{RP} (a) shows time variation of the dark matter density profile. At 1 Gyr, the dark matter follows the initial density profile given in Eq. (\ref{pseudo}). After the collapse in the first 1 Gyr, the shape of the dark matter profile does not change over 10 Gyr.\\
\indent Figure \ref{RP}(b) shows time variation of the gas density profile. Inner region ($\lesssim$ 1 kpc) of the gas density profile follows the evolution of the dark matter density profile. Outer region of gas is blown away due to the outflow induced by SNe. In addition, the total amount of gas reduces because of star formation. However, gas still remains even at 10 Gyr. As in the Appendix, all of our models have gas at 10 Gyr. The observed LG dSphs in contrast have no or little gas \citep[e.g.][]{McConnachie12}. This result suggests that physical processes such as ram pressure and tidal stripping are required to remove all gas away from dSphs \citep{Mayer06, Nichols14}. \\
\indent As shown in Figure \ref{RP} (c), the stellar density profile of our simulation well reproduces observations. In Figure \ref{RP} (c), we present the stellar density profiles. Stars distribute within $\sim$ 1 kpc, which is consistent with the observed tidal radii ($\sim$0.5 -- 3 kpc) of dSphs in the LG \citep{Irwin95}. The density profile of stars is basically associated with the dark matter density profile.\\
\indent Figure \ref{RP} (d) shows the stellar velocity dispersion profile. The observed stellar velocity dispersion of dSphs is almost constant within $\sim$ 1 kpc from the center \citep{Walker09}. Model s000 has similar properties with the observed radial stellar velocity dispersion profiles inside 1 kpc from the center in the LG dSphs.
\begin{figure}[htbp]
\epsscale{1.2}
\plotone{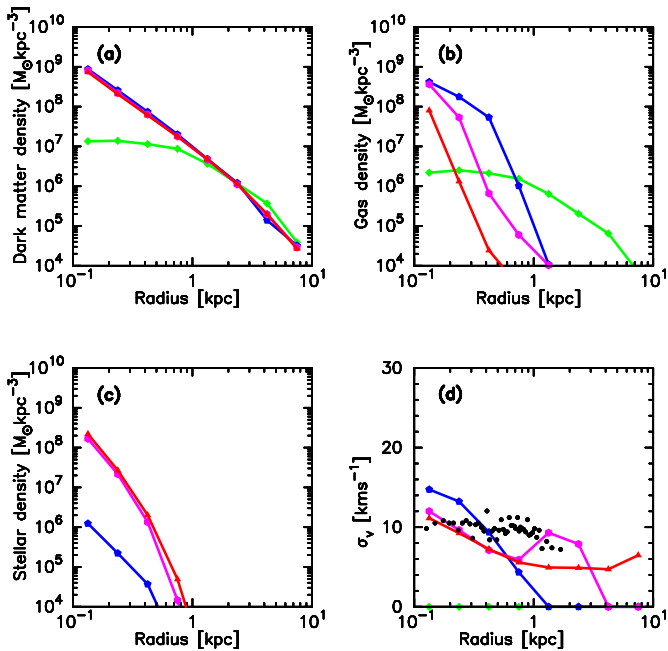}
\caption{Radial profiles of model s000 at $t$ = 0 Gyr (green), 1 Gyr (blue), 5 Gyr (magenta), and 10 Gyr (red). (a): radial dark matter density profile. (b): radial gas density profile. (c): radial stellar density profile. (d): radial stellar velocity dispersion profile. Black dots are observed stellar velocity dispersion in the Fornax dSph \citep{Walker09}. \label{RP}}
\end{figure}
\subsection{Time variations of the star formation rate}\label{Star_Formation}
\begin{figure}[htbp]
\epsscale{1.5}
\plotone{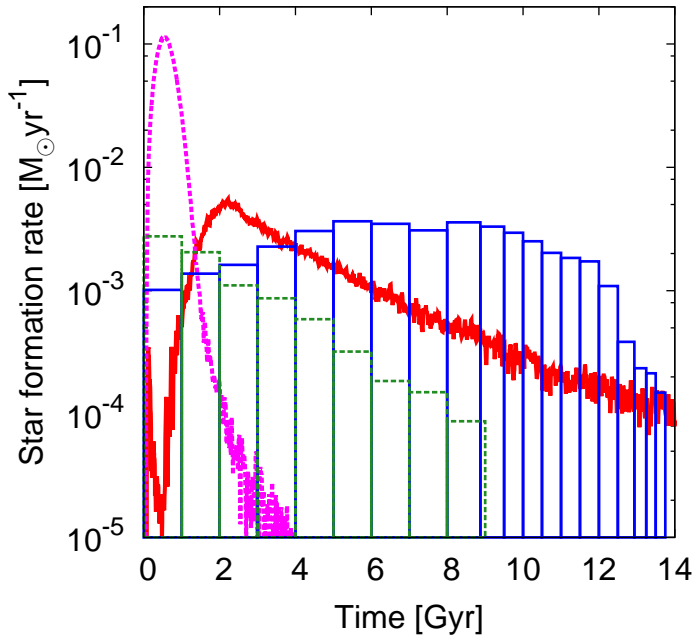}
\caption{The SFR as a function of time for our models. Red and magenta-dashed curves represent the SFR of models s000 and mExt, respectively. Blue and green-dashed histograms represent the observed SFR of the Fornax dSph \citep{deBoer12} and the Sculptor dSph \citep{deBoer12a}, respectively. \label{SFR}}
\end{figure}

\indent Figure \ref{SFR} shows the time variation of the SFR in model s000 (red curve) and the observed values of the Fornax and the Sculptor dSphs \citep{deBoer12a, deBoer12}. The SFR of model s000 is peaked at $\sim$ 2 Gyr. Gas density increases with accretion (see Figure \ref{RP} (b)) and finally reaches the threshold density for star formation. On the other hand, SN feedback drives gas away from the inner region to the outer region \citep{Hopkins11}. Because of the shallow gravitational potential and high threshold density for star formation ($n_{\mathrm{th}}$ = 100 cm$^{-3}$), SN feedback significantly affects the timescale of gas accretion. It therefore takes long time ($\sim$ 1 Gyr) to reach the peak of the SFR. The SFR of model s000 ($\sim 10^{-3} M_{\odot}$ yr$^{-1}$) is consistent with the observed value of the Fornax and the Sculptor dSphs inferred from color-magnitude diagram analysis \citep[$\sim 10^{-3} M_{\odot}$ yr$^{-1}$,][]{deBoer12a, deBoer12}. The SFR of model mExt (magenta curve) is discussed in \S\ref{p_m}. 
\subsection{Metallicity distribution}\label{Metallicity Distribution}
\begin{figure}[htbp]
\epsscale{1.2}
\plotone{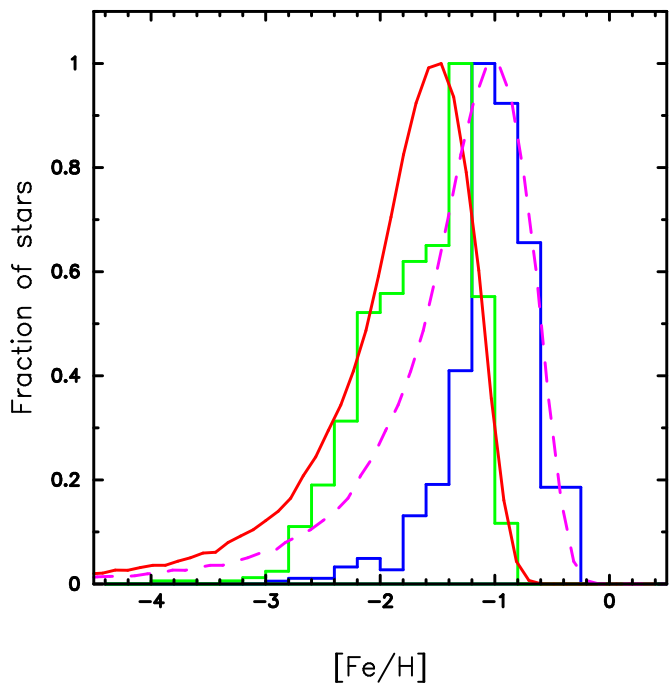}
\caption{Metallicity distribution of model s000 (red curve), model s000 but [Fe/H] is shifted to 0.5 dex taking into account the effect of SNe Ia (magenta-dashed curve), the observed value of the Fornax dSph (blue histogram) \citep{Kirby10}, and the Sculptor dSph (green histogram) \citep{Kirby09, Kirby10, Kirby12}. \label{MD}}
\end{figure}

Metallicity distribution is one of the best properties to test reliability of chemical evolution models. Figure \ref{MD} compares metallicity distribution between model s000 and observation. All data of the Fornax \citep{Kirby10} and the Sculptor dSphs \citep{Kirby09, Kirby10, Kirby12}. Metallicity distribution of model s000 is almost consistent with the observed value of the Sculptor dSph. The metallicity at the peak of the distribution of model s000 is [Fe/H] = $-$1.57, which is lower than that of the Fornax dSph, [Fe/H] = $-$1.06 \citep{Kirby13}.  This is because we do not implement SNe Ia in this model, while the Fornax dSph must be significantly affected by the metal ejection of SNe Ia \citep[e.g., ][]{Kirby10}. If we take account of the products of SNe Ia, the peak metallicity is expected to shift by $\sim$ 0.5 dex to the higher metallicity, which is closer value to that of the Fornax dSph.\\ 
\section{Enrichment of $r$-process elements in dwarf spheroidal galaxies}\label{$r$-process}
\subsection{Enrichment of $r$-process elements without metal mixing in star-forming region}\label{mix1}
\begin{figure}[htbp]
\epsscale{1.2}
\plotone{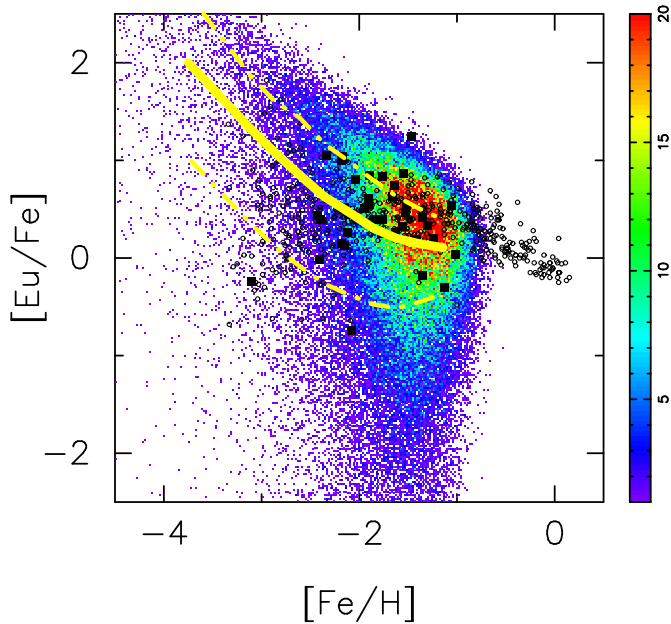}
\caption{[Eu/Fe] as a function of [Fe/H] of model s000. Contour is the number of stars produced in our model, between 0 (purple) and 20 (red). Yellow curve is the median of model prediction. Dash-dotted curves are the first and third quartiles, respectively. Circles are the observed value of the Galactic halo stars \citep[SAGA database, ][]{Suda08, Suda11, Yamada13}. Squares are the observed value of stars in Carina, Draco, Leo I, Sculptor, and Ursa Minor dSphs \citep[SAGA database,][]{Suda14}. Stars produced by our model are plotted within 0.5 kpc from the center of our model galaxies. \label{rnomix}}
\end{figure}

\begin{figure}[htbp]
\epsscale{2.4}
\plottwo{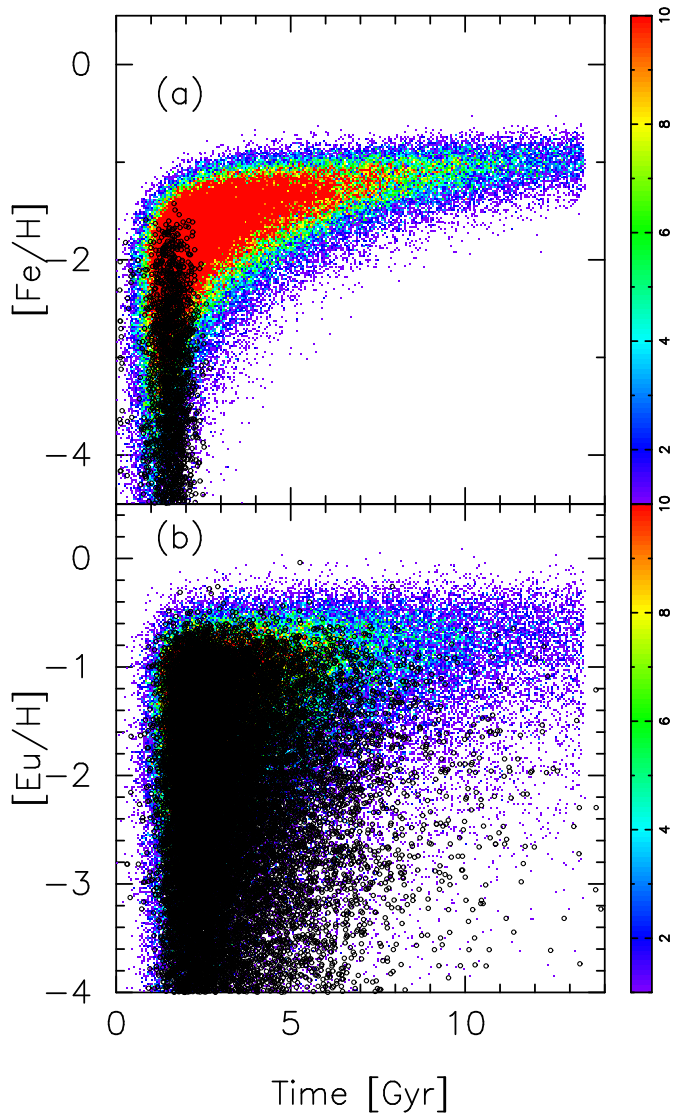}{f6b.eps}
\caption{(a): [Fe/H] as a function of time from the start of the simulation of model s000. (b): [Eu/H] as a function of time from the start of the simulation of model s000. Black circles are stars formed from gas particles, which are affected by one CCSN. Contour is the same as Figure \ref{rnomix}. \label{AgeFe}}
\end{figure}

In this section, we discuss [Eu/Fe] as a function of [Fe/H] predicted in model s000. In model s000, the metallicity of a star particle inherits that of the gas particle from which the star particle was formed, according to the method (1) in \S\ref{metal-stars}. Figure \ref{rnomix} shows [Eu/Fe] as a function of [Fe/H] predicted in model s000. We also put the observed data of the Galactic halo and several dSphs, i.e., Carina, Draco, Leo I, Sculptor, and Ursa Minor dSphs \citep[SAGA database, ][]{Suda08, Suda11, Suda14, Yamada13}, excluding carbon-enhanced stars, which are possibly affected by gas transfer in binaries. We also exclude stars in the Fornax dSph because some of them have extremely high [Eu/Fe] ($>$ 0.5 dex) due to significant contamination of $s$-process \citep{Letarte10}. As shown in Figure \ref{rnomix}, stars of highly $r$-process enhanced stars; [Eu/Fe] $>$ 1 (so-called $r$-II stars), are over-abundant. In addition, $r$-deficient stars in $-2 < $ [Fe/H] $< -1$ are predicted. Such low [Eu/Fe] stars are not seen in the observation. These stars are not simply caused by delayed production of Eu by NSMs. In fact, the average value of [Eu/Fe] does not increase with metallicity at around [Fe/H] $\sim -2$. The significant dispersions of chemical components among gas particles seem to be rather essential. The relations between the galactic age and abundances of Fe and Eu show the reason why such large dispersions are seen in low metallicity region. \\
Figure \ref{AgeFe} (a) shows [Fe/H] as a function of time. Metallicity obviously increases with time, as CCSNe produce iron. The metallicity however has more scatters at the earlier time, especially during the first few Gyrs. Then, later formed stars are enriched by more numerous CCSNe, and as a result, the dispersion of stellar metallicity decreases with time. We denote the stars, which are formed from the gas enriched only by a single CCSN by black circles in Figure \ref{AgeFe} (a). Their metallicity widely distributes over $\sim$ 3 dex. These stars concentrate only in $\lesssim$ 2 Gyr. \\
\indent Figure \ref{AgeFe} (b) shows [Eu/H] as a function of time. In contrast to Figure \ref{AgeFe} (a), large star-to-star scatters in [Eu/H] remain over the whole evolution of the galaxy. As shown in black circles in Figure \ref{AgeFe} (b), gas particles affected by one NSM remain over 10 Gyr. One of the reasons must be the low rate of NSMs. The rate of NSMs is one hundred times lower than that of CCSNe in this model. The total number of NSMs may not be enough to converge the [Eu/H] in this model.\\
In addition to the low NSM rate, the efficiency of gas mixing in this model seems to cause unnaturally large scatters in [Eu/H].
If a star particle contains products from a single NSM, the value of [Eu/H] must be determined by the distance from the NSM, which enriched the star-forming gas particle. However, as shown in Figure \ref{AgeFe} (b), dispersion of such stars drags longer than 5 -- 10 Gyrs, which is much longer than the merger time of NSMs. It implies that gas particles never change the abundance of Eu, unless other NSMs enrich them again, though in actual, gas clouds are expected to interact with others.\\
\indent In fact, observations of open clusters show that stellar metallicity is quite homogeneous in each cluster \citep[e.g., ][]{De Silva07a, De Silva07b, Pancino10, Bubar10, De Silva11, Ting12, Reddy12, De Silva13, Reddy13}. The gas in star-forming region is possibly homogenized by hydro-dynamical effects, such as turbulent mixing \citep{Feng14}.\\
\indent In addition, previous studies suggest that the standard SPH simulations without metal mixing tend to predict a lower amount of gas with low metallicity and higher metallicity of inter galactic medium \citep{Wiersma09, Shen10}. \citet{Shen15} suggest that it is difficult to reproduce the observed [Eu/Fe] as a function of [Fe/H] without metal mixing in star-forming region.\\
\indent On the other hand, the fiducial model of \citet{Voort15} reproduces the observed [Eu/Fe] of metal-poor stars, although they adopt the same definition of metallicity as our model s000. They suggest that large-scale metal mixing such as galactic winds and hydrodynamical flows is important to reproduce the observed [Eu/Fe] as a function of [Fe/H]. In their model, NSMs eject metals in the region of $3.5\times10^{6} M_{\odot}$, which is much larger than the swept-up mass of NSMs ($\sim 10^{4} M_{\odot}$). The treatment of \citet{Voort15} is identical to implement metal mixing.
\subsection{Effects of metal mixing on enrichment of $r$-process elements}\label{mix2}
\begin{figure}[htbp]
\epsscale{1.2}
\plotone{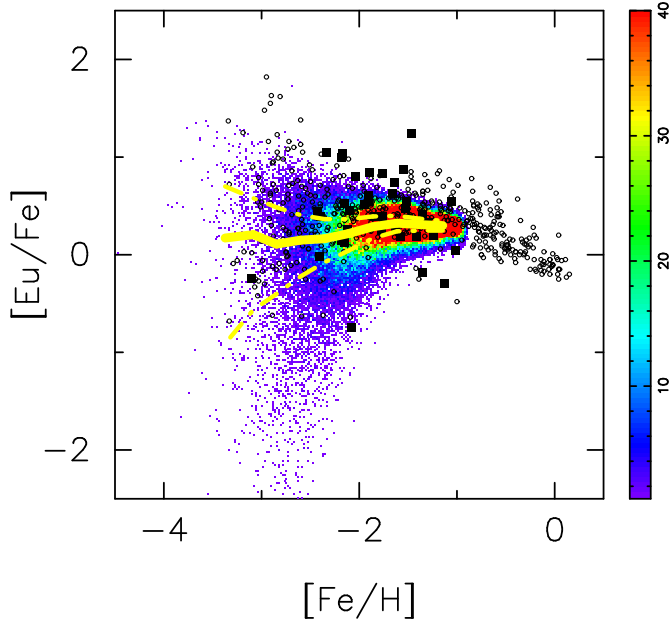}
\caption{[Eu/Fe] as a function of [Fe/H] of the model m000. Symbols are the same as Figure \ref{rnomix}.\label{rmix}}
\end{figure}

\begin{figure}[htbp]
\epsscale{1.2}
\plotone{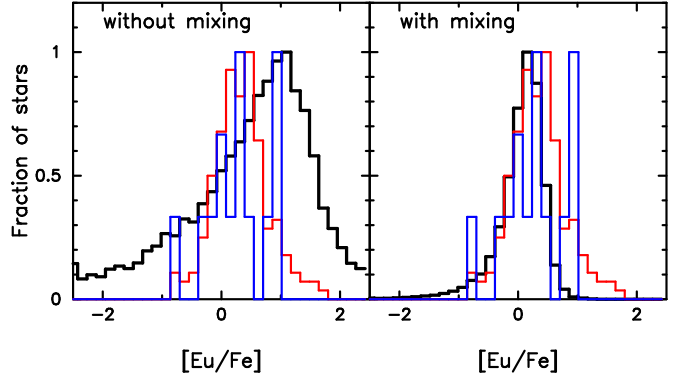}
\caption{[Eu/Fe] distribution of stars with [Fe/H] $< -2.0$ of our models (black histogram) and observation (red histogram: the Galactic halo stars, blue histogram: Carina, Draco, Leo I, Sculptor, and Ursa Minor dSphs). Data are compiled by SAGA database \citep{Suda08, Suda11, Suda14, Yamada13}. Left panel: plotted model is s000 (without metal mixing in star-forming region). Right panel: plotted model is m000 (with metal mixing in star-forming region).\label{Eu_Fe_mix}}
\end{figure}

\begin{figure}[htbp]
\epsscale{2.4}
\plottwo{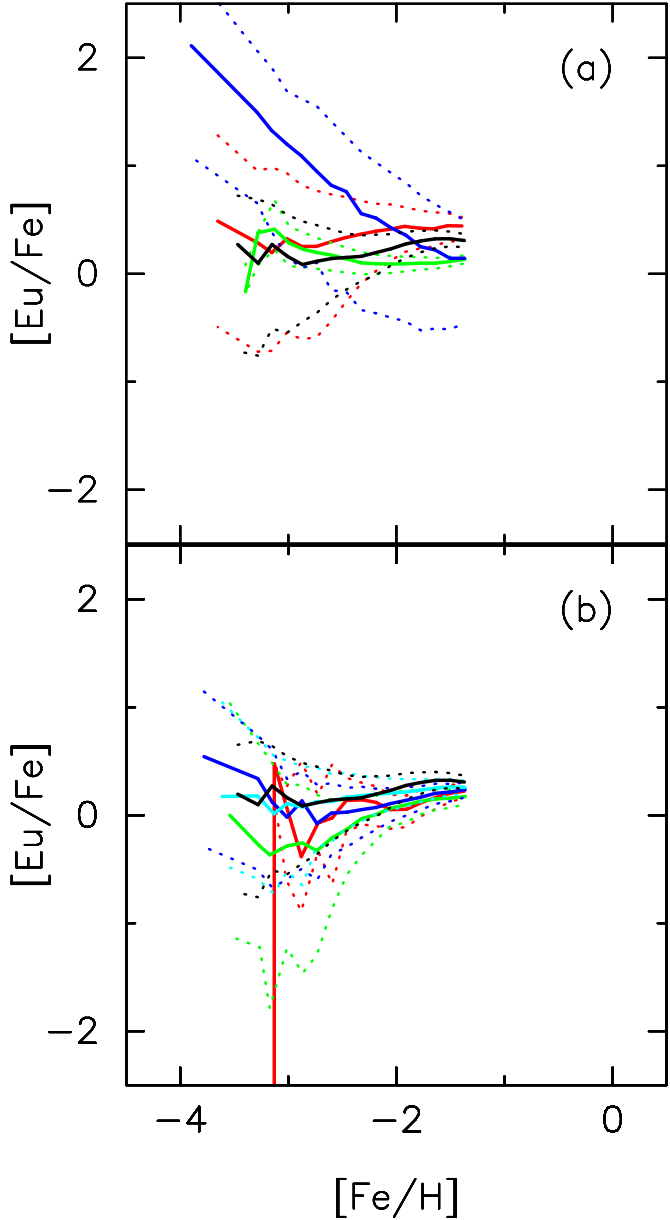}{f9b.eps}
\caption{(a) Median [Eu/Fe] as a function of [Fe/H] produced by the different number of nearest neighbor particles ($N_{\mathrm{ngb}}$). Solid curves are the median of model prediction and dashed curves are the first and third quartiles. Red curve represents mN16 ($N_{\rm ngb}$ = 16). Black curve represents m000 ($N_{\rm ngb}$ = 32). Green curve represents mN64 ($N_{\rm ngb}$ = 64). Blue curve represents s000 (without metal mixing in star-forming region). (b) Median [Eu/Fe] as a function of [Fe/H] produced by different initial number of particles ($N$). Red curve represents m014 ($N = 2^{14}$). Green curve represents m016 ($N = 2^{16}$).  Blue curve represents m017 ($N = 2^{17}$). Sky blue curve represents m018 ($N = 2^{18}$). Black curve represents m000 ($N = 2^{19}$).\label{convergence}}
\end{figure}

As discussed in the previous section, the effect of mixing of enriched gas must be essential to account for the observed values of [Eu/Fe] in metal-poor stars. Therefore, we take account of the effect of metal mixing according to the method (2) in \S\ref{metal-stars}, and we adopt the average metallicity of gas particles in the SPH kernel of the progenitor gas particle for the metallicity of newly formed stars. Figure \ref{rmix} shows [Eu/Fe] as a function of [Fe/H] with the metal mixing model, m000. Model m000 has the same parameter set for s000 except for the effect of metal mixing in star-forming region (see Table \ref{list}). As shown in this figure, $r$-deficient stars in [Fe/H] of $-2$ to $-1$ seen in Figure \ref{rnomix} disappear due to metal mixing. The fraction of $r$-II stars is also reduced in this model due to adopted averaged metallicity in star-forming region. Model m000 apparently reproduces the observational tendency of [Eu/Fe] in metal-poor stars much better than the model s000. Our model does not require the assumption of short merger time ($t_{\mathrm{NSM}} \lesssim$ 10 Myr), which is required to reproduce observations in previous studies \citep[e.g., ][]{Argast04, Matteucci14, Komiya14, Tsujimoto14}.  
\mbox{}\\
\indent Figure \ref{Eu_Fe_mix} shows [Eu/Fe] distributions in stars of  [Fe/H] $< -2.0$ predicted in s000 (without mixing model) and m000 (with mixing model). Observational values of the MW (red histogram) and dSphs (blue histogram) are provided by the SAGA database \citep{Suda08, Suda11, Suda14, Yamada13}. While model s000 overproduces stars with [Eu/Fe] $<-1$, model m000 significantly reduces the fraction of $r$-deficient stars. In addition, the fraction of $r$-II stars also reduces in model m000. This result therefore suggests that metal mixing in star-forming region is fairly important physical process to reproduce the observed [Eu/Fe] as a function of [Fe/H].\\
\indent The predicted distribution of [Eu/Fe] must be affected by the mass of the mixed gas ($M_{\mathrm{mix}}$), and the initial total number of particles ($N$). We define $M_{\rm mix} = N_{\rm ngb}m_{\rm gas}$. The mass of one gas particle ($m_{\rm gas}$) is proportional to $N^{-1}$. Table \ref{mixing_region} lists the examined values of parameters and the corresponding mixing mass ($M_{\mathrm{mix}}$). Figure \ref{convergence} (a) shows the median value and the dispersion of [Eu/Fe] of models, which have different $N_{\rm ngb}$. The dispersion of [Eu/Fe] becomes smaller with larger $N_{\mathrm{ngb}}$ as shown in this figure. When we increase the $N_{\mathrm{ngb}}$, the value of $M_{\mathrm{mix}}$ increases and larger mass fraction of gas in the galaxy is mixed. Due to the effects of increasing $M_{\mathrm{mix}}$, the dispersion becomes smaller in the models with larger value of $N_{\mathrm{ngb}}$. \\
\indent Figure \ref{convergence} (b) shows the median value of [Eu/Fe] as a function of [Fe/H] of models, which have different $N$. As shown in Figure \ref{convergence} (b), the dispersion of [Eu/Fe] shows the similar tendency irrespective of $N$, i.e., it decreases with increasing of metallicity. When we adopt the larger value of $N$, more star particles are produced, i.e., the number of events of metal mixing increases. This means that metals are more mixed in the star-forming region if $M_{\rm mix}$ is constant. However, $M_{\mathrm{mix}}$ is defined to be proportional to $N^{-1}$. Models with larger value of $N$ have smaller value of $M_{\rm mix}$. The effect of increasing $N$ offsets the effect made by decreasing $M_{\rm mix}$. Thus, the dispersion is not affected by $N$. 
\begin{deluxetable}{lrrr}
\tabletypesize{\scriptsize}
\tablecaption{Mass of metal mixing region. \label{mixing_region}}
\tablewidth{0pt}
\tablehead{
\colhead{Model}&
\colhead{$N$}&
\colhead{$N_{\mathrm{ngb}}$}&
\colhead{$M_{\mathrm{mix}}$}\\
	&&&\colhead{($10^{4} M_{\odot}$)}}
\startdata
m000&2$^{19}$&32&1.3\\
mN16&2$^{19}$&16&0.6\\
mN64&2$^{19}$&64&2.7\\
m018&2$^{18}$&32&2.6\\
m017&2$^{17}$&32&5.1\\
m016&2$^{16}$&32&10.3\\
m014&2$^{14}$&32&41.0
\enddata
\tablecomments{The columns correspond to the name of model, initial total number of particles ($N$), the number of nearest neighbor particles ($N_{\mathrm{ngb}}$), and mass of the mixing region ($M_{\mathrm{mix}}$). }
\end{deluxetable}
\mbox{}\\
\subsection{The relationship between the relative $r$-process abundance ratio and the star formation rate}\label{p_m}
\begin{figure}[htbp]
\epsscale{1.2}
\plotone{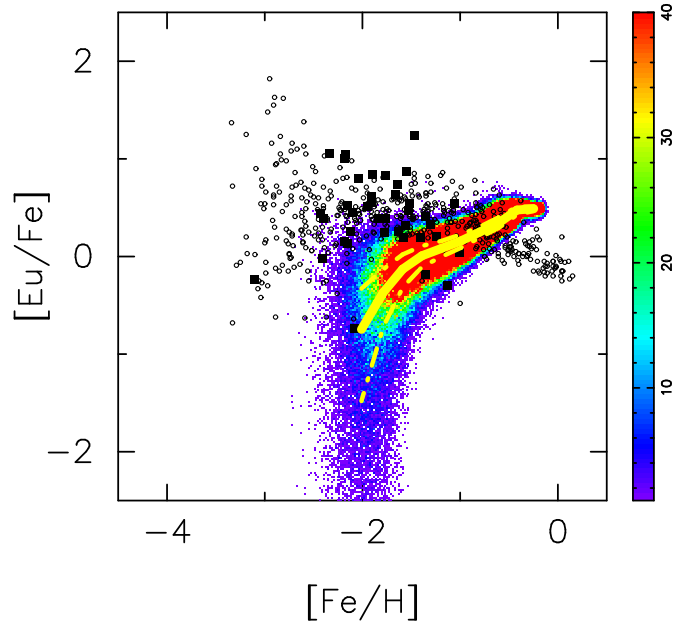}
\caption{[Eu/Fe] as a function of [Fe/H] of model mExt ($M_{\mathrm{tot}}$ = $7\times$10$^{8}$ $M_{\odot}$, $c_{\star}$ = 0.033, $n_{\mathrm{th}}$ = 0.1 cm$^{-3}$, $\epsilon_{\mathrm{SN}}$ = 3$\times$10$^{49}$ erg, and $t$ = 100 Myr). Symbols are the same as Figure \ref{rnomix}. \label{FHRm}}
\end{figure}

The value and scatters in the abundance ratio of [Eu/Fe] must be affected by SFR, especially for metal-poor stars.
In order to examine the effect of SFR, 
we discuss [Eu/Fe] as a function of [Fe/H] with the model mExt, in which extremely weak SN feedback energy ($\epsilon_{\mathrm{SN}} = 3\times10^{49}$ erg) and low threshold density for star formation ($n_{\mathrm{th}}$ = 0.1 cm$^{-3}$) are assumed (Table \ref{list}).
According to Appendix \ref{app}, the weaker feedback model produces higher SFR and model with smaller value of $n_{\mathrm{th}}$ shifts the peak of SFR to the earlier time. Since mExt has lower values of $\epsilon_{\mathrm{SN}}$ and $n_{\mathrm{th}}$, stars are more easily produced comparing to the model m000. The SFR of model mExt  during the first 1 Gyr rises rapidly and is larger than $10^{-2} M_{\odot}$yr$^{-1}$. The peak of SFR reaches $\sim10^{-1}M_{\odot}$yr$^{-1}$, while that of m000  is $\lesssim10^{-3} M_{\odot}$yr$^{-1}$. 
The SFR of model mExt at $\lesssim$ 1 Gyr is 
much higher than the observational values (Figure \ref{SFR}).\\
\indent Figure \ref{FHRm} shows [Eu/Fe] as a function of [Fe/H] in model mExt. This model predicts significantly different [Eu/Fe] from m000 (Figure \ref{rmix}) although both models adopt the same initial distribution of gas particles. Model mExt produces a distribution of $r$-deficient stars around [Fe/H] $\sim -$2. In addition, no stars are produced below [Fe/H] $\sim -$3.\\
\indent This difference is related to the timescale of chemical evolution in the early phase of the galaxy formation. For model m000, the median metallicity of gas particles at 1 Gyr is [Fe/H] = $-$3.32. On the other hand, the median metallicity of gas particles in mExt is [Fe/H] $= -$0.91 at 1 Gyr. The SFE of model mExt is estimated to be $\sim 0.1$--1 Gyr$^{-1}$. The value of SFE is comparable to some other inhomogeneous chemical evolution studies \citep{Argast04, Cescutti15, Wehmeyer15}.
Model mExt proceeds chemical evolution much faster than m000 due to the high SFR of mExt ($\sim 10^{-2} M_{\odot}$yr$^{-1}$). The [Eu/Fe] as a function of [Fe/H] in model mExt is inconsistent with the observation due to fast chemical evolution by the high SFR. This result suggests that the SFR and SFE in the early phase of dSphs are $\lesssim 10^{-3} M_{\odot}$yr$^{-1}$ and $\lesssim$ 0.10 Gyr$^{-1}$, respectively, if the $r$-process elements are ejected by NSMs with a long merger time ($\sim$100 Myr). As shown in Figure \ref{SFR}, the SFR of $\sim 10^{-3} M_{\odot}$yr$^{-1}$ is a reasonable agreement with the observed value of the Fornax dSph \citep{deBoer12}. This SFR is also consistent with sub-halo models of \citet{Ishimaru14} (Case 1 in their model suggests that
the appropriate value of star formation efficiency for a sub-halo
with a stellar mass of $10^7 M_\odot$ should be 0.10 Gyr$^{-1}$,
which corresponds to the order of SFR as $\sim 10^{-3} M_{\odot}$yr$^{-1}$). They also suggest that the observed [Eu/Fe] scatter in metal-poor stars by NSMs with a long merger time. 
\subsection{Merger time of neutron star mergers}\label{tdel}
\begin{figure}[htbp]
\epsscale{2.4}
\plottwo{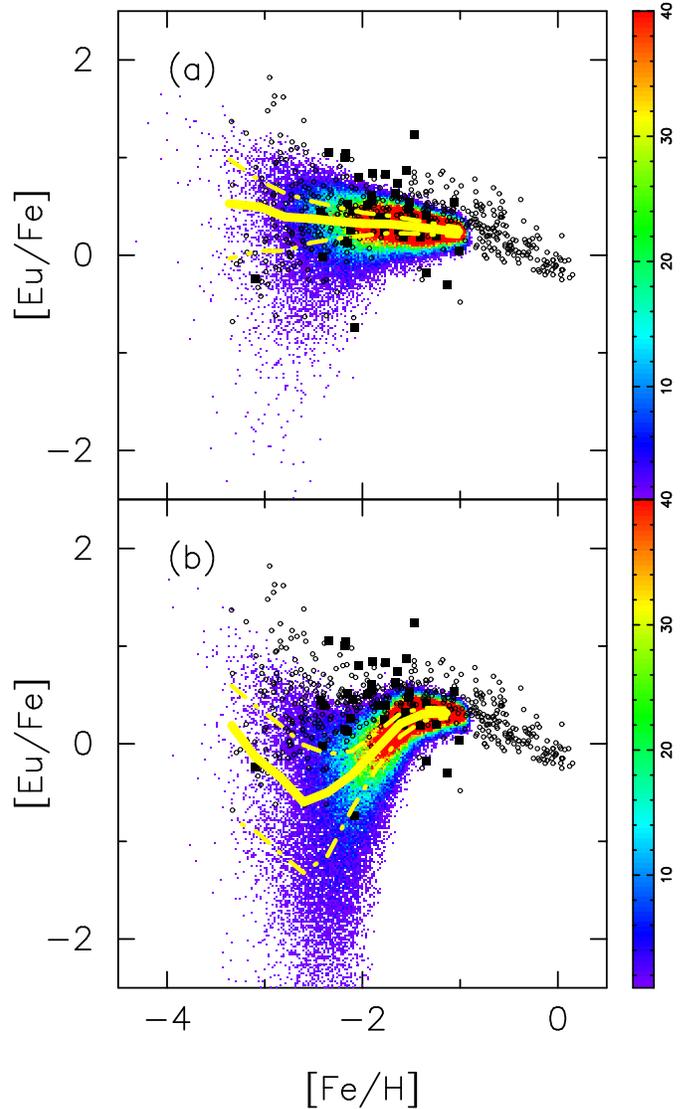}{f11b.eps}
\caption{[Eu/Fe] as a function of [Fe/H] with different merger time of NSMs. (a): mt10 ($t_{\mathrm{NSM}}$ = 10 Myr). (b): mt500 ($t_{\mathrm{NSM}}$ = 500 Myr). Symbols are the same as Figure \ref{rnomix}.\label{delay}}
\end{figure}

\begin{figure}[htbp]
\epsscale{1.2}
\plotone{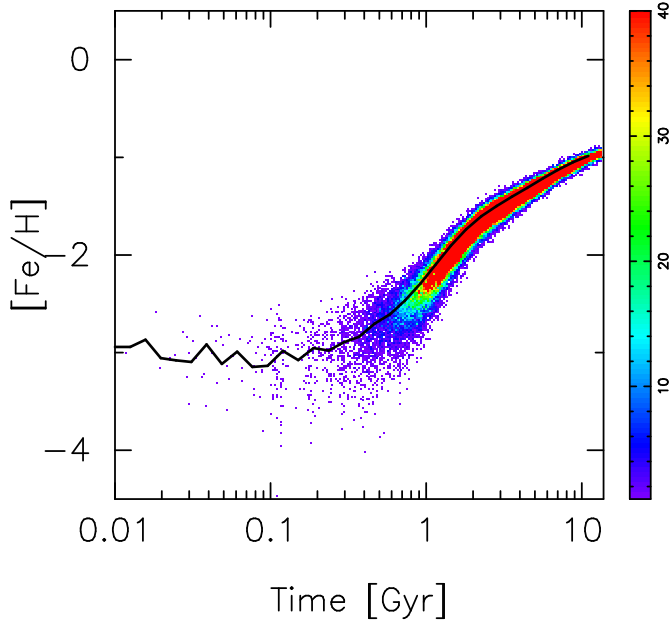}
\caption{[Fe/H] as a function of time of model m000. The horizontal axis is plotted from 600 Myr from the start of the simulation. Black curve is the average of the metallicity in each age. Contour is the same as Figure \ref{rnomix}. \label{Age_log}}
\end{figure}

In this section, we discuss the effect of merger time of NSMs. Figure \ref{delay} shows resultant [Eu/Fe] as a function of [Fe/H] by NSMs with different merger time ($t_{\mathrm{NSM}}$). Eu in Figure \ref{delay} (a) and (b) are produced by NSMs with $t_{\mathrm{NSM}}$ = 10 Myr (mt10) and 500 Myr (mt500), respectively. Although mt10 has a slightly smaller fraction of stars in $-3 <$ [Fe/H] $<-2$ than model m000, the global relative abundance ratio is similar to m000 ($t_{\mathrm{NSM}}$ = 100 Myr). Contrary to the models m000 and mt10, the model with much longer merger time such as 500 Myr in mt500 shows large scatters in [Eu/Fe] at higher metallicity and cannot account for the observed scatters in [Fe/H]$\sim -3$.\\
\indent Figure \ref{Age_log} shows [Fe/H] as a function of the substantial galactic age, i.e., the elapsed time from the rise of the major star formation. As shown in Figure 3, we can regard that the major star formation arises from 600 Myr from the beginning of the calculation. The average metallicity of stars is almost constant during the first $\sim$ 300 Myr. Due to low star formation efficiency of the galaxy, spatial distribution of metallicity is highly inhomogeneous in $\lesssim$ 300 Myr. In this epoch, since most of gas particles are enriched only by a single SN, metallicity of stars is mainly determined simply by the distance from each SN to the gas particles which formed the stars. Therefore, NSMs with $t_{\mathrm{NSM}} \sim 100$ Myr can account for the observation of EMP stars, as well as those with $t_{\mathrm{NSM}} \sim 10$ Myr. In contrast, metallicity is well correlated with the galactic age after $\sim$ 300 Myr, irrespective of the distance from each SN to the gas particles. Because SN products have already been well mixed in a galaxy, the stellar metallicity is determined by the number of the SNe, which enriched the stellar ingredients. Therefore, if the merger time of NSMs is much longer than $\sim$ 300 Myr, it is too long to reproduce observations.\\
\subsection{The rate of neutron star mergers}\label{NSMrate}
\begin{figure}[htbp]
\epsscale{2.4}
\plottwo{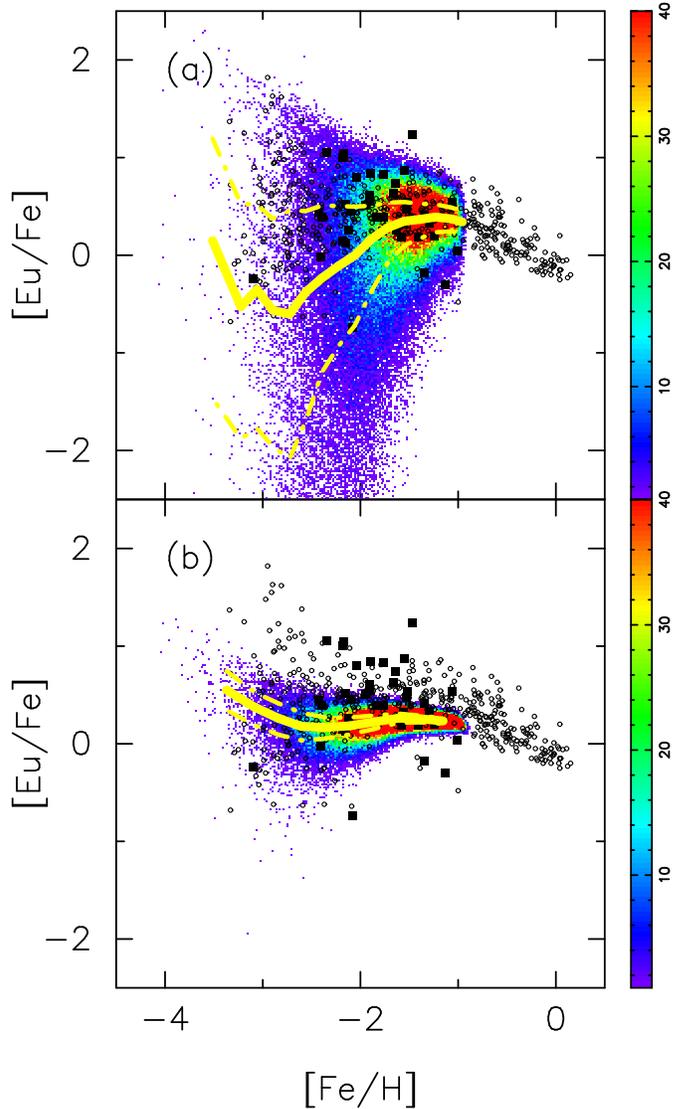}{f13b.eps}
\caption{[Eu/Fe] as a function of time with different rate of NSMs. (a): mr0.001 ($f_{\mathrm{NSM}}$ = 0.001). (b): mr0.1 ($f_{\mathrm{NSM}}$ = 0.1). Symbols are the same as Figure \ref{rnomix}. \label{rate}}
\end{figure}

The yields of $r$-process elements in our models are related to the NSM rate as already mentioned in \S\ref{Chem}, 
though the
Galactic rate of NSMs 
is highly uncertain. The estimated Galactic
NSM rate is 10$^{-6}$ to 10$^{-3}$ yr$^{-1}$  based on three observed
binary pulsars \citep{Abadie10}. Table
\ref{yield} lists yields of models discussed here. Figure \ref{rate}
shows predicted [Eu/Fe] as a function of [Fe/H] assuming different NSM
rate. Figure \ref{rate} (a) and (b) represent models with the NSM fractions
$f_{\mathrm{NSM}}$ = 0.001 (mr0.001) and $f_{\mathrm{NSM}}$ = 0.1
(mr0.1), respectively. 
The corresponding NSM rate in a MW-like galaxy is
$\sim 10^{-5}$ yr$^{-1}$ (mr0.001) and $\sim 10^{-3}$ yr$^{-1}$ (mr0.1). 
Model mr0.001 predicts larger scatter and a smaller number of stars at [Fe/H] $<-$3 than m000. Model mr0.001 has [Eu/Fe] dispersion by more than 3 dex at
[Fe/H] = $-$2. In addition, there remains $\sim$ 1 dex dispersion even for stars with [Fe/H] $>-$2. 
In contrast, model mr0.1 predicts smaller scatter than m000, though it does not seem to be inconsistent with observations. Such tendencies are also seen in \citet{Argast04}, \citet{Komiya14} and
\citet{Voort15}.\\ 
\indent Our fiducial model, m000, reproduces the observed $r$-process
ratio as discussed in \S\ref{mix2}. The NSM rate of m000 for a MW-like galaxy
is $\sim 10^{-4}$ yr$^{-1}$. The total mass of $r$-process elements produced by each NSM corresponds to
$\sim 10^{-2} M_{\odot}$. The value is consistent with recent nucleosynthesis calculations: $10^{-3} M_{\odot}$ to $10^{-2} M_{\odot}$ \citep[e.g., ][]{Goriely11, Korobkin12, Hotokezaka13, Bauswein13, Wanajo14}.\\
\indent \citet{Argast04} construct an inhomogeneous chemical evolution model of the MW halo. Their model is difficult to reproduce [Eu/Fe] by NSMs with the Galactic NSM rate of $2\times 10^{-4}$ yr$^{-1}$ due to high star formation efficiency. [Eu/Fe] produced in their model is similar to that of mExt (Figure \ref{FHRm}). \\
\indent From the discussion above, NSM rate of $\sim$ 10$^{-4}$ yr$^{-1}$ in a MW size galaxy is preferred to reproduce the observed [Eu/Fe]. This rate is consistent with the estimated galactic NSM rate from the observed binary pulsars \citep{Abadie10}. Near future gravitational detectors, KAGRA, advanced LIGO, and advanced VIRGO  \citep{Abadie10b, Kuroda10, Accadia11, LIGO13} are expected to detect 10 -- 100 events per year of gravitational wave from NSMs.\\
\begin{deluxetable}{lll}
\tabletypesize{\scriptsize}
\tablecaption{List of yields. \label{yield}}
\tablewidth{0pt}
\tablehead{
\colhead{Model}&
\colhead{$f_{\mathrm{NSM}}$}&
\colhead{$M_{\mathrm{r}}$}\\
	&&\colhead{($M_{\odot}$})}
\startdata
mr0.001&0.001&10$^{-1}$\\
m000&0.01&10$^{-2}$\\
mr0.1&0.1&10$^{-3}$
\enddata
\tablecomments{The columns correspond to the name of models, fraction of NSMs ($f_{\mathrm{NSM}}$), and total yields of $r$-process elements ($M_{\mathrm{r}}$). Fraction of NSMs is the fraction of stars that cause NSMs in the mass range 8 -- 20 $M_{\odot}$.}
\end{deluxetable}
\section{Summary}
We have carried out numerical simulations of chemo-dynamical evolution of dSphs using $N$-body/SPH code, ASURA to investigate the enrichment history of $r$-process elements.  This study suggests that NSMs with the merger time of $\sim$ 100 Myr and the Galactic event rate of $\sim 10^{-4}$ yr$^{-1}$ can explain the dispersion of [Eu/Fe] in reasonable agreement with observations in EMP stars. This study supports that NSMs are the major astrophysical site of the $r$-process. Our isolated dSph models reproduce basic properties of the observed LG dSphs such as radial profiles, time variations of the SFR as well as metallicity distribution. Here, we summarize the main results. 
\begin{itemize}
\item[(1)]{The abundance ratio of [Eu/Fe] produced in our models without metal mixing in star-forming regions has too large dispersion. This is because metals in a gas particle do not diffuse out to mix into the other particles throughout the evolution of galaxies.}
\item[(2)]{Models with metal mixing in star-forming region reproduce the observed [Eu/Fe] distribution and its scatter as a function of [Fe/H]. Our model shows good convergence of the resolution. We show that NSMs with $t_{\mathrm{NSM}}$ = 100 Myr is favorable for reproducing the observed [Eu/Fe] as a function of metallicity. This result implies that the metal mixing process is critical to reproduce the [Eu/Fe] distribution. In addition, this study suggests that the SFR of dSphs in the early epoch of their evolution $\sim 1$Gyr is $\lesssim 10^{-3} M_{\odot}$yr$^{-1}$.}
\item[(3)]{The NSMs with merger time of $\lesssim$ 300 Myr is acceptable to account for the observed abundance of EMP stars. This is because metallicity is not correlated with time up to $\sim$ 300 Myr from the start of the simulation due to low star formation efficiency of the model galaxy.}
\item[(4)]{This study suggests that the Galactic NSM rate to account for the observed $r$-process abundance scatters is $\sim 10^{-4}$ yr$^{-1}$.  The total mass of $r$-process elements ejected by one NSM is $\sim 10^{-2} M_{\odot}$, which is consistent with the value suggested by nucleosynthesis studies ($10^{-3}$--$10^{-2} M_{\odot}$). Next generation gravitational detectors KAGRA, advanced LIGO and advanced VIRGO are expected to detect gravitational wave from NSMs and their event rate would be over 10 per year. Their detections will give us more reliable galactic NSM rate.}
\end{itemize}

In this study, we have focused on the enrichment history of $r$-process elements in isolated dSphs with fixed mass. To fully understand the enrichment history of $r$-process elements in the LG galaxies, it is important to show how the mass and size of galaxies affect the enrichment of $r$-process elements.

Recent observations suggest that the low abundance of $r$-process elements ([Ba/Fe]\footnote{Barium (Ba) can also be regarded as the $r$-process element in [Fe/H] $\lesssim -3$} $<-1$) in stars with [Fe/H] $<-3.5$. These stars provide clues to understand the astrophysical site(s) of $r$-process elements and the metal enrichment in the first galaxies. Some studies suggest that the $r$-process abundance of these stars can be explained by short merger time channel ($\sim$ 1 Myr) of NSMs \citep{Ishimaru14} or accretion of materials from inter stellar medium to the Population III stars \citep{Komiya14}. Since we have only focused on the star-to-star scatters of [Eu/Fe] in stars with [Fe/H] $\sim -$3 assuming that NSMs are the major site of $r$-process in this paper, we have not discussed the origin of these stars. It is profitable to discuss the origin of these stars by the detailed simulation of galaxies.


To address all of these issues, it is required to understand how the MW halo formed. We need to clarify the relation between the building block galaxies of the MW and the present LG dSphs. It is now underway to extend our numerical simulations of chemodynamical evolution of dSphs to larger scale simulations of the MW in order to fully understand the enrichment of $r$-process elements in the MW and confirm the validity of the scenario of hierarchical structure formation.
\acknowledgments
We appreciate the referee for giving us profitable comments to improve our paper. We are grateful to Masaomi Tanaka, Shinya Wanajo, Grant J. Mathews, and Yuichiro Sekiguchi for fruitful discussion. We also appreciate for Takuma Suda for providing us the latest version of SAGA database including the dataset of the LG dSphs. This work is supported in part by MEXT SPIRE and JICFuS. YH is financially supported by Grant-in-Aid for JSPS Fellows (15J00548). The authors are financially supported by JSPS Grants-in-Aid for Scientific Research (YI: 26400232, TRS: 26707007, MSF: 26800108, TK: 26105517, 24340060, 15H03665). Numerical computations and analysis were in part carried out on Cray XC30 and computers at Center for Computational Astrophysics, National Astronomical Observatory of Japan.
\appendix
\section{Parameter dependence}
\subsection{Radial profile}
\begin{figure}[htbp]
\epsscale{1.2}
\plotone{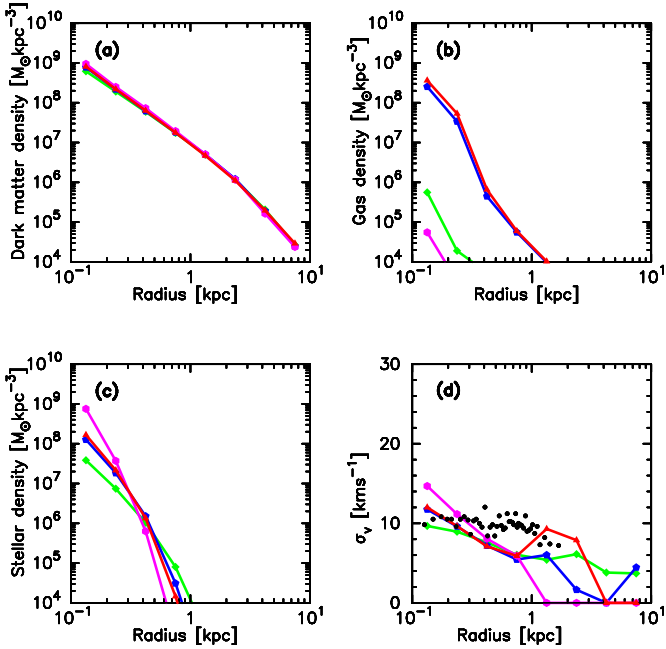}
\caption{Radial profiles of models of different parameters at $t$ = 5.0 Gyr. Red triangles, green diamonds, blue pentagons, and magenta hexagons represent model s000, sn10 ($n_{\mathrm{th}}$ = 10 cm$^{-3}$), c050s ($c_{\star}$ = 0.5), and se01 ($\epsilon_{\mathrm{SN}} = 10^{50}$ erg), respectively. (a): Radial dark matter density profile. (b): Radial gas density profile. (c): Radial stellar density profile. (d): Radial velocity dispersion profile.\label{RP_appendix}}
\end{figure}

We compare models with different threshold densities for star formation ($n_{\mathrm{th}}$), dimensionless SFE parameters ($c_{\star}$), and SN feedback energies ($\epsilon_{\mathrm{SN}}$). Table \ref{list_ap} lists all models discussed here. Figure \ref{RP_appendix} shows the radial profiles of our models. Horizontal axis of Figure \ref{RP_appendix} is the distance from the galactic center. Figure \ref{RP_appendix} (a) shows the dark matter density profile. We find that all models have similar dark matter profiles. The dark matter profile is not affected by physical parameters such as threshold density for star formation, dimensionless SFE parameters, and SN feedback energy.\\
\indent Gas density, stellar density, and stellar velocity dispersion profiles have variations among these models. The gas density of sn10 and se01 is lower than that of s000 (Figure \ref{RP_appendix} (b)). In these models, most of gas is consumed in the early phase of their evolution. The gas density of se01 is truncated at 0.3 kpc while the gas profiles of the other models continue over 10 kpc. This is because the feedback energy in se01 is too weak to blow the gas away to the outer region of the galaxy. \\
\indent Stellar density profiles for all models (Figure \ref{RP_appendix} (c)) are truncated within a few kpc, which is consistent with the observed truncation radius ($\sim$0.5 -- 3 kpc) of dSphs in the LG \citep{Irwin95}. In sn10, stars distribute to a larger radius than the other models. When a low value of $n_{\mathrm{th}}$ is used, stars can form in the outer region of the galaxy. If the SN feedback is weak (model se01), stellar distribution concentrates on the central region of the galaxy.\\
\indent Figure \ref{RP_appendix} (d) shows the velocity dispersion profile. All models except for se01 are consistent with the observed radial velocity dispersion profiles inside 1 kpc from the center in the LG dSphs \citep{Walker07, Walker09}. Model se01 has higher velocity dispersion at the center of our model galaxy due to the high central concentration of stars. \\
\indent In contrast to $n_{\mathrm{th}}$ and $\epsilon_{\mathrm{SN}}$, the dimensionless SFE parameter ($c_{\star}$) does not greatly affect the radial profiles. Figure \ref{RP_appendix} shows all profiles of s000 (red curve) and sc50 (blue curve) are similar although they have different value of the dimensionless SFE parameter. These features suggest that radial properties of galaxies are insensitive to the value of $c_{\star}$ when we adopt a reasonable value of $n_{\mathrm{th}}$ (= 100 cm$^{-3}$) \citep{Saitoh08}. \\
\begin{deluxetable}{lrll}
\tabletypesize{\scriptsize}
\tablecaption{List of models. \label{list_ap}}
\tablewidth{0pt}
\tablehead{
\colhead{Model}&
\colhead{$n_{\mathrm{th}}$}&
\colhead{$c_{\star}$}&
\colhead{$\epsilon_{\mathrm{SN}}$}\\
	&\colhead{cm$^{-3}$}&&\colhead{10$^{51}$ erg}}
\startdata
s000&100&0.033&1\\sn10&10&0.033&1\\sc50&100&0.5&1\\se01&100&0.033&0.1
\enddata
\tablecomments{Parameters adopted in our models: (1) Model: Name of our models. (2) $c_{\star}$: Dimensionless star formation efficiency parameter. (3) $n_{\mathrm{th}}$: Threshold density for star formation. (4) $\epsilon_{\mathrm{SN}}$: SN feedback energy.}
\end{deluxetable}
\subsection{Time variations of the star formation rate and metallicity distribution}\label{app}
\begin{figure}[htbp]
\epsscale{1.5}
\plotone{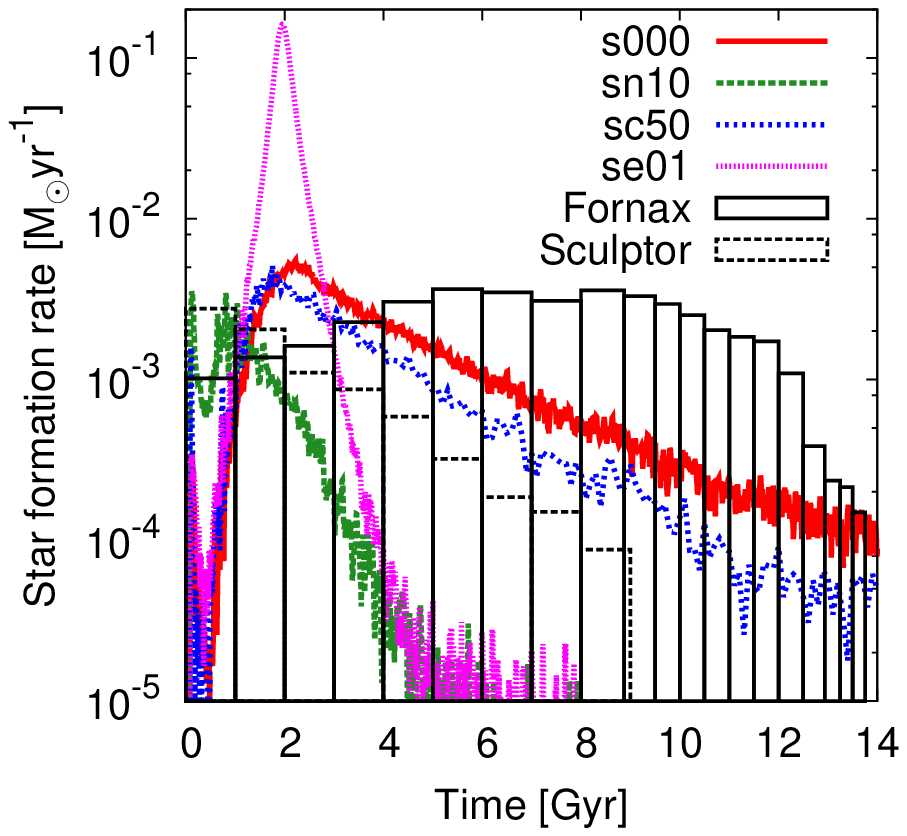}
\caption{The SFR as a function of time for our models. Red curve, green dashed curve, blue short-dashed curve, and magenta dotted curve represent s000, sn10 ($n_{\mathrm{th}}$ = 10 cm$^{-3}$), sc50 ($c_{\star}$ = 0.5), and se01 ($\epsilon_{\mathrm{SN}} = 10^{50}$ erg), respectively.  The black histogram and black-dotted histogram represent the observed SFR of the Fornax dSph \citep{deBoer12} and the Sculptor dSph \citep{deBoer12a}, respectively. \label{SFR_appendix}}
\end{figure}

\begin{figure}[htbp]
\epsscale{1.2}
\plotone{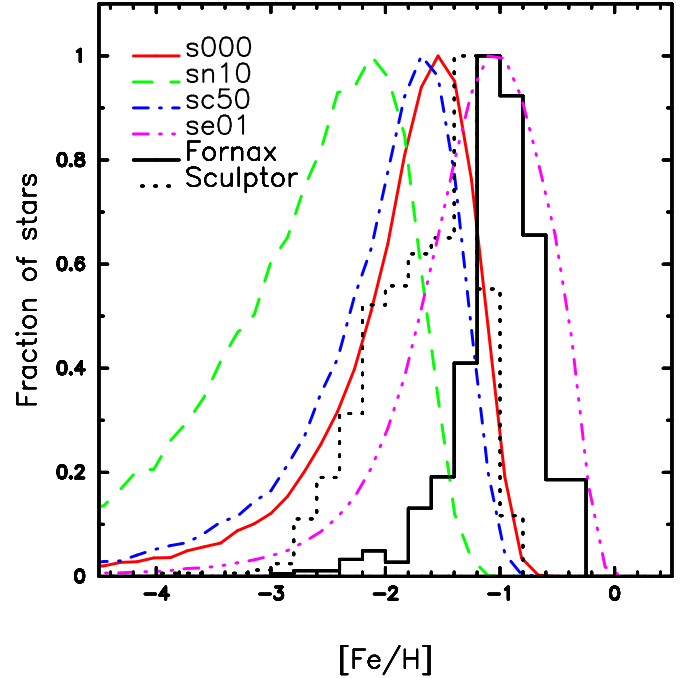}
\caption{Metallicity distribution of our models. The black histogram and black-dotted histogram are observed metallicity distribution of the Fornax dSph \citep{Kirby10} and the Sculptor dSph \citep{Kirby09, Kirby10, Kirby12}, respectively. Red curve, green dashed curve, blue dash-dotted curve, and magenta dash-dot-dotted curve represent s000, sn10 ($n_{\mathrm{th}}$ = 10 cm$^{-3}$), sc50 ($c_{\star}$ = 0.5), and se01 ($\epsilon_{\mathrm{SN}} = 10^{50}$ erg), respectively. \label{MD_appendix}}
\end{figure}

\indent Figure \ref{SFR_appendix} and \ref{MD_appendix} show SFR as a function of time and metallicity distribution, respectively. SFR as a function of time and metallicity distribution are characterized by the threshold density for star formation and SN feedback energy. Model sn10 has lower $n_{\mathrm{th}}$ (= 10 cm$^{-3}$) than that of s000 ($n_{\mathrm{th}}$ = 100 cm$^{-3}$). The second peak of SFR of sn10 is earlier than s000. This reflects time required to reach $n_{\mathrm{th}}$ is shorter than s000 because of using low $n_{\mathrm{th}}$ value in sn10. In addition, the first peak of SFR of sn10 is higher than s000. Gas is consumed by star formation and removed by outflow at $<$ 0.1 Gyr in sn10. Its SFR is therefore $\sim$2 dex lower than s000. The early bursty star formation of sn10 produces more metal-poor stars than s000 (see green dashed and red curves in Figure \ref{MD_appendix}). Higher SFR of sn10  $\sim$ 0.1 Gyr produces more CCSNe in this phase. CCSNe produce outflow. Model sn10 thus loses larger amount of gas around $\sim$ 0.1 Gyr. Chemical evolution of sn10, therefore, quenches at $> $ 0.1 Gyr and produces only few metal-rich stars. \\
\indent SN feedback energy also significantly affects the time variation of SFR and metallicity distribution. The SN feedback energy of se01 ($\epsilon_{\mathrm{SN}} = 10^{50}$ erg) is ten times smaller than that of s000 ($\epsilon_{\mathrm{SN}} = 10^{51}$ erg). The peak of the SFR of se01 is over 1 dex higher than that of s000. SN feedback energy gives thermal energy to gas particles. It prevents collapse of gas particles. As a result, star formation is suppressed due to SN feedback energy. SN feedback energy of se01 is too weak to suppress star formation. The peak of SFR of se01 is thus $\sim$ 1 dex higher than s000. Gas is consumed for star formation due to high SFR around 2 Gyr of se01, and the SFR at $ >$ 4 Gyr is eventually suppressed. Due to the low SN feedback energy in model se01, the peak of metallicity distribution of se01 is $\sim$ 0.5 dex higher metallicity than s000. \\
\indent On the other hand, the value of the dimensionless SFE parameter does not significantly affect the results. Models s000 and sc50 have different value of $c_{\star}$ = 0.033 and 0.5, respectively. The time variation of the SFR and metallicity distribution of sc50 is however similar to that of s000. This result suggests that the time variation of the SFR and metallicity distribution are insensitive to the value of $c_{\star}$ if we use a reasonable value of $n_{\mathrm{th}}$  (= 100 cm$^{-3}$).  Slightly lower metallicity of sc50 than s000 is due to slightly lower SFR of sc50 than s000. This result suggests that the value of $c_{\star}$ does not affect the metallicity distribution as well as radial profiles and the SFR. When we adopt $n_{\mathrm{th}}$ = 100 cm$^{-3}$, it takes much longer local dynamical time to flow from the reservoir ($n_{\mathrm{H}}\sim$ 1 cm$^{-3}$) to the star-forming regions ($n_{\mathrm{H}} \gtrsim$ 100 cm$^{-3}$). This timescale does not depend on $c_{\star}$ \citep{Saitoh08}. Our results are, thus, independent of $c_{\star}$.\\ 
\indent These results suggest that the time variation of the SFR and metallicity distribution is significantly affected by the threshold density for star formation and SN feedback energy. Low $n_{\mathrm{th}}$ model (sn10) produces too many EMP stars. In contrast, low $\epsilon_{\mathrm{SN}}$ model (se01) has too many metal-rich stars. These differences in metallicity distribution are due to the difference of the time variation of the SFR among models. On the other hand, model s000 reproduces observation of metallicity distribution as well as dynamical properties. Parameters of s000 are taken from the observed values. Threshold density of s000 ($n_{\mathrm{th}}$ = 100 cm$^{-3}$) is taken from mean density of GMCs. SN feedback energy of s000 ($\epsilon_{\mathrm{SN}}$ = $10^{51}$ erg) is taken from the canonical explosion energy of CCSNe \citep[e.g.,][]{Nomoto06}. We thus treat s000 as a model that has fiducial parameter sets.\newpage
\mbox{}\\\newpage

\clearpage
\end{document}